\documentclass[12pt]{article}
\usepackage{subeqn,amsfonts}

\newfont{\twelvemsb}{msbm10 scaled\magstep1}
\newfont{\eightmsb}{msbm8}
\newfam\msbfam
\textfont\msbfam=\twelvemsb
\scriptfont\msbfam=\eightmsb
\catcode`\@=11
\def\Bbb{\ifmmode\let\next\Bbb@\else
  \def\next{\errmessage{Use \string\Bbb\space only in math mode}}\fi\next}
\def\Bbb@#1{{\fam\msbfam{{#1}}}}

\newcommand{\be}{\begin{equation}}
\newcommand{\ee}{\end{equation}}
\newcommand{\ba}{\begin{eqnarray}}
\newcommand{\ea}{\end{eqnarray}}

\newcommand{\NP}[1]{Nucl.\ Phys.\ {\bf #1}}
\newcommand{\PL}[1]{Phys.\ Lett.\ {\bf #1}}

\newcommand{\CMP}[1]{Comm.\ Math.\ Phys.\ {\bf #1}}

\newcommand{\MPL}[1]{Mod.\ Phys.\ Lett.\ {\bf #1}}

\begin{document}
\sloppy
\renewcommand{\thefootnote}{\fnsymbol{footnote}}

\newpage
\setcounter{page}{1}

\vspace{0.7cm}
\begin{flushright}
DCPT-03/05\\
EMPG-02-21\\
April 2003
\end{flushright}
\vspace*{1cm}
\begin{center}
{\bf  Exact conserved quantities on the cylinder II: off-critical case.}\\
\vspace{1.8cm}
{\large D.\ Fioravanti $^a$ and M.\ Rossi $^b$ \footnote{E-mail:
Davide.Fioravanti@durham.ac.uk, M.Rossi@ma.hw.ac.uk}}\\ \vspace{.5cm}
$^a${\em Department of Mathematical Sciences,
South Road, Durham DH1 3LE,             
England} \\

\vspace{.3cm}
$^b${\em Department of Mathematics, Heriot-Watt University, Edinburgh EH14 4AS,
Scotland} \\ 
\end{center}
\vspace{1cm}

\renewcommand{\thefootnote}{\arabic{footnote}}
\setcounter{footnote}{0}

\begin{abstract}
{\noindent With} the aim of exploring a massive model corresponding to the perturbation of the conformal model \cite {FR2} the nonlinear integral equation for a quantum system
consisting of left and right KdV equations coupled on the cylinder
is derived from an integrable lattice field theory. The eigenvalues
of the energy and of the transfer matrix (and of all the other local integrals of motion) are expressed in
terms of the corresponding solutions of the nonlinear integral
equation. The analytic and asymptotic behaviours of the transfer matrix are studied and given. 
\end{abstract}

\vspace{1cm}


\newpage

\section {Introduction}

The computation of the eigenvalues of the conserved charges - in particular the local ones - is a very
important issue in the study of $1+1$ dimensional integrable
models. For instance, the local integrals of motion have recently revealed to enter the form of off-critical null-vectors \cite {BBS}. Among the various methods of computing, a possible one uses a
discretisation of the space and the diagonalisation of the discrete
transfer matrix by means of a Bethe Ansatz technique.
The problem of recovering the continuous limit of such eigenvalues can be
tackled using nonlinear integral equations, method previously applied to genuine lattice models like the XXZ chain and the inhomogeneous six-vertex model \cite {KP,DDV,FMQR}.

In a previous paper \cite {FR} we coupled on the lattice the Lax
operators of two - left and right - quantum (m)KdV
equations, proposing two slightly different monodromy matrices as
possible descriptions of perturbed minimal conformal field theories. The Bethe
Ansatz diagonalization of the discrete transfer matrices was
completely achieved, showing that the problem is somehow equivalent to a
twisted spin $-1/2$ XXZ chain with alternating
dishomogeneities. The twist was called {\it dynamical}, because it is
automatically generated by the theory (and depends on the number of Bethe roots), not introduced as an external ad hoc
parameter. 
We concluded that this lattice model describes, in some sense, a dynamically twisted version of the lattice sine-Gordon theory \cite {IK} and, in this respect, we conjectured and partially
argued that in the continuous limit it gives a description of perturbed minimal conformal models.   

In this letter we address the problem of performing the continuous
limit of the eigenvalues of the aforementioned transfer matrices \cite {FR}, by
finding the nonlinear integral equation obeyed by the counting function. In particular, exact expressions of all the local integrals of motion are given as nonlinear functionals of the counting function. This
problem is different from that solved in \cite {DDV,FMQR}, since these papers used a lattice approach to the twisted 
sine-Gordon model based on an inhomogeneous spin $+1/2$ chain (with a suitably added twist). However, this
lattice discretisation does not furnish a transfer matrix for the
sine-Gordon model, whereas our
construction is endowed with an
off-critical transfer matrix, which generates all the integrals of motion, in particular the local ones. On the other hand, our lattice model satisfies a braided version of the Yang-Baxter algebra \cite {HK}.

\section {The nonlinear integral equation for the vacuum sector}

In this paper we continue the work done in \cite {FR2}, extending it to the massive case in an unifying view of conformal and off-critical theory. In fact, we start by recalling the introduction of the mass \cite {FR}, realised through coupling two KdV theories via dishomogeneities. More precisely, we considered on a lattice of length
$R$, spacing $\Delta$ and number of sites $N=R/\Delta$, $N \in 4 {\mathbb N}$, 
the following monodromy matrices 
\ba  {\bf M }(\alpha)&\equiv&e^{-\frac {i\pi \beta ^2}{2}}\prod
_{i=1}^{\stackrel {N/4}{\leftarrow}}\bar L_{4i}(\alpha +\Theta )\bar L_{4i-1}(\alpha+\Theta)L_{4i-2}(\alpha-\Theta)L_{4i-3}(\alpha-\Theta) \,, \label {Rightmon} \\ 
{\bf M}^\prime (\alpha)&\equiv&e^{\frac {i\pi \beta ^2}{2}}\prod
_{i=1}^{\stackrel {N/4}{\leftarrow}}L_{4i}(\alpha
-\Theta )L_{4i-1}(\alpha-\Theta)\bar L_{4i-2}(\alpha+\Theta)\bar
L_{4i-3}(\alpha+\Theta)  \,
\label {Leftmon} \, ,
\ea
with coupling constant $0\leq \beta ^2 \leq 1$ and dishomogeneity $\Theta$ 
real. In these formul{\ae} left Lax operators $L_m(\alpha)$ and right Lax operators
$\bar L_m(\alpha)$ are given by
\begin{equation}
L_{m}(\alpha) \equiv \left (
\begin{array}{cc} e^{-iV_m^-} & e^{\alpha}  e^{iV_m^+} \\ e^{\alpha}
e^{-iV_m^+}&  e^{iV_m^-}\\ \end{array} \right ) \, ,\quad  \bar L_{m}(\alpha)
\equiv \left ( \begin{array}{cc} e^{-iV_m^+} &  e^{-\alpha}   e^{iV_m^-}\\
e^{-\alpha}  e^{-iV_m^-} &   e^{iV_m^+}\\  \end{array} \right ) \, ,
\label{lax}
\end{equation}
where $V^{\pm}_m$ are the discretised quantum counterparts of the mKdV
variables $v$ (left case) and $\bar v$ (right case) and satisfy the
nonultralocal commutation relations
\begin{eqnarray} &&[V^{\pm}_m\, , \,
V^{\pm}_n]=\pm \frac {i\pi \beta ^2}{2}(\delta _{m-1,n}-\delta _{m,n-1})
\, , \label {vrel1} \\ 
&&[V^+_m\, , \,
V^-_n]=-\frac {i\pi \beta ^2}{2}(\delta _{m-1,n}-2\delta
_{m,n}+\delta _{m,n-1}) \, , \label {vrel2} 
\end{eqnarray}
and the periodicity conditions
$V^{\pm}_{m+N}=V^{\pm}_m$. In fact, we have proved in \cite {FR} that,
as a consequence of (\ref {vrel1}, \ref {vrel2}), monodromy matrices (\ref
{Rightmon}) and (\ref {Leftmon}) satisfy braided Yang-Baxter 
equations \cite {HK}. Moreover, the diagonalisation of the corresponding 
transfer matrices \cite {FR} is based on the Bethe equations 
\begin{equation}
e^{-\varepsilon2i\pi \beta ^2 l}\prod _{{\stackrel{r=1}{r\not= s}}}^l \frac {\sinh (\alpha _s -\alpha _r +i\pi \beta ^2)}{\sinh (\alpha _s -\alpha _r -i\pi \beta ^2)}=\left [\frac
{\sinh \left (\alpha _s +\Theta -\frac {i\pi \beta ^2}{2}\right)\sinh \left(\alpha _s -\Theta -\frac {i\pi\beta ^2}{2}\right)}
{\sinh \left (\alpha _s +\Theta +\frac {i\pi \beta ^2}{2}\right)\sinh
\left (\alpha _s -\Theta +\frac {i\pi \beta ^2}{2}\right)}
\right]^{N/4} \, ,
\label{Bethe}
\end{equation}
in which $l$
is the number of Bethe roots and $\varepsilon $ is equal to
$+1$ for (\ref {Rightmon}) and to $-1$ for (\ref
{Leftmon}). Correspondingly, the eigenvalues on Bethe states of the transfer matrices are
\be
{\bf \Lambda}_N(\alpha )={\bf \Lambda}^+_N(\alpha )+{\bf \Lambda}^-_N(\alpha ) \, ,
\label {tra} 
\ee
where
\be
{\bf \Lambda}^\pm_N(\alpha )=  
e^{\mp \varepsilon i\pi \beta ^2 l} \prod\limits _{r=1}^{l} \frac {\sinh (\alpha -\alpha _r \pm i\pi \beta ^2)}{\sinh (\alpha -\alpha _r) } \rho _N^{\pm}(\alpha)
\,  \label{tra0} 
\ee
and
\be
\rho _N ^\pm (\alpha )= 
e^{-\frac {\Theta N}{2}} \left [ 4 \sinh \left ( \Theta -\alpha 
\mp \frac {i\pi \beta ^2}{2} \right )
\sinh \left (\Theta +\alpha 
\pm \frac {i\pi \beta ^2}{2} \right ) \right ]^{N/4}\, .
\label {rho} 
\ee
Comparison with equation (2.18) of \cite {EFIK} shows that, after the
identifications $2\cosh 2\Theta=1/S$, $\alpha
_s=\lambda _s -i\pi /2$ and the substitution $\beta ^2 \rightarrow \beta ^2 /8\pi$, Bethe equations (\ref {Bethe}) coincide with
the Bethe equations for the lattice sine-Gordon model \cite {IK}  
twisted by the dynamical factor $e^{-\varepsilon 2i\pi \beta ^2 l}$.

Now we want to summarise all the Bethe equations (\ref {Bethe}) into a single nonlinear integral equation. As a preliminary,
we define the function, analytic in the strip $|{\mbox {Im}}\, x| <\,
{\mbox {min}}\, \{\zeta, \pi -\zeta\} \, , \, \, 0\leq \zeta < \pi$, 
\begin{equation}
\phi (x,\zeta)\equiv i\ln \frac {{\mbox {sinh}} (i\zeta+x)}{{\mbox {sinh}} (i\zeta-x)}\, , \label {funz}
\end{equation}
which allows us to define the {\it counting function}
\begin{equation}
Z_N(x)\equiv \frac {N}{4}\left [\phi \left (x +\Theta, \frac {\pi \beta ^2}{2} \right )+\phi \left(x -\Theta, \frac {\pi \beta ^2}{2}\right) \right]
+\sum _{r=1}^{l}\phi (x -\alpha _r, \pi \beta ^2)+2\pi \beta ^2 \varepsilon l  \, . \label{count0}
\end{equation}
In terms of the counting function the equations (\ref {Bethe}) assume the form
\begin{equation}
Z_N(\alpha _s)=\pi (2I_s-l-1) \, , \quad   I_s \in {\mathbb N} \, . \label {cond1}
\end{equation}
Between the various solutions to the Bethe equations we single out the
vacuum, i.e. the eigenstate with the minimum energy, for which $\alpha _s$ are real and $Z_N(\alpha _s)$ are equal
to all the numbers of the form (\ref {cond1}) between $Z_N(-\infty)$ and
$Z_N(+\infty)$. From (\ref {count0}) it follows that these two extremes are
\be
Z_N(\pm \infty)=\pm \frac {N}{4}(2\pi-2\pi \beta ^2)\pm l(\pi-2\pi \beta ^2)+2\pi \beta ^2 \varepsilon l \, .\label {extr}
\ee
The vacuum minimises the energy - which we will define in next section - when
\be
1/3<\beta ^2 <2/3 \, , \label {range}
\ee
and the ``twist'' $2\pi \beta ^2 \varepsilon l$ is small enough mod. $2\pi$.
Considering $\beta ^2$ in the region (\ref {range}), we now want to write an equation for the counting function associated to the vacuum in the continuous limit: $N\rightarrow +\infty$, $R$ fixed.
We firstly remark that (\ref {extr}) imply that in the continuous
limit also $l\rightarrow +\infty$ and 
therefore, even if we define
\be
\beta ^2 =p/p^\prime \, , \quad p<p^\prime \quad {\mbox
  {coprimes}}\, , 
\ee
the last term in (\ref {count0}) is
divergent. Therefore, we stick to the different subsequences
\be
l=np^\prime+\kappa \quad , \quad 0\leq \kappa \leq p^\prime -1 \, , \label {par1}
\end{equation}
where $n \rightarrow +\infty$ with $\kappa $ fixed.
Discarding in the definition of $Z_N(x)$ multiples of $2\pi$ we get:
\be
Z_N(x)=\frac {N}{4}\left [\phi \left (x +\Theta, \frac {\pi}{2}\beta ^2 \right )+\phi \left(x -\Theta, \frac {\pi}{2}\beta ^2\right) \right]+
\sum _{r=1}^{l}\phi \left (x -\alpha _r, \pi \beta ^2 \right )+2\pi \varepsilon \omega  \, , \label{count1}
\ee
where $\omega$ is finite in the continuous limit (the double braces denote the fractional part):
\be
\omega = \left \{ \left \{ \frac {p\kappa}{p^\prime}\right \} \right \} \, . \label{omega}
\ee
Now, using standard techniques (see \cite {DDV}), one can
rewrite a sum over the vacuum Bethe roots of a function $f_N$ with no poles on a strip around 
the real axis as follows: \footnote {We are fixing $l$ to be even in
  the following (i.e. $e^{iZ_N(\alpha _s)}=-1$), as the case $l$ odd
  can be treated along the same lines and in the end does not add new information.} 

\be
\sum _{r=1}^{l}f_N(\alpha _r)=-\int _{-\infty}^{+\infty}\frac {dx}{2\pi} f^\prime _N(x)Z_N(x)+2\int _{-\infty}^{+\infty}\frac {dx}{2\pi}f^\prime_N (x){\mbox {Im}}\ln \left [1+e^{iZ_N(x+i0)}\right ] \, . \label {sumf1}
\ee
Applying (\ref {sumf1}) to the sum contained in (\ref {count1}) we get
\begin{eqnarray}
&&Z_N(x)=\frac {N}{4}\left[\phi \left(x +\Theta, \frac {\pi}{2}\beta ^2\right)+
\phi \left(x -\Theta, \frac {\pi}{2}\beta ^2\right)\right]+2\pi
\varepsilon \omega + \label{count2} \\
&+&\int _{-\infty}^{+\infty}\frac {dy}{2\pi} \phi ^\prime (x-y, \pi \beta ^2)Z_N(y)-2\int _{-\infty}^{+\infty}\frac {dy}{2\pi}\phi ^\prime (x-y,\pi \beta ^2){\mbox {Im}}\ln \left [1+e^{iZ_N(y+i0)}\right ] \nonumber
\, . 
\end{eqnarray}
Let us define for sake of conciseness
\begin{eqnarray}
&K(x)\equiv\frac {1}{2\pi}\phi ^{\prime}\left (x, \pi \beta ^2\right )\, , \quad \Phi (x)\equiv   \phi \left(x +\Theta, \frac {\pi}{2}\beta ^2\right)+
\phi \left(x -\Theta, \frac {\pi}{2}\beta ^2\right)\, , & \nonumber \\
&L_N(x)\equiv{\mbox {Im}}\ln \left [1+e^{iZ_N(x+i0)}\right ]\, . &  \label {not}
\end{eqnarray}
In these notations the Fourier transform of relation (\ref {count2}) is
\be
\hat Z_N(k)=\frac {N}{4}\frac {\hat \Phi (k)}{1-\hat K(k)}-2\frac {\hat K(k)}{1-\hat K(k)}
\hat L_N(k)+4\pi ^2 \varepsilon \omega 
\frac  {\delta (k)}{1-\hat K(k)} \, , \label {count3}
\ee
where we have adopted the convention
$\hat f(k)\equiv \int _{-\infty}^{+\infty}dx e^{-ikx} f(x)$
about the Fourier transform.
Explicitly, the Fourier transforms of the first two functions in (\ref {not}) are respectively
\be
\hat K(k)=\frac {\sinh k\left (\frac {\pi}{2}-\pi \beta
    ^2\right)}{\sinh \frac {\pi}{2}k} \, , \quad \hat \Phi (k)= \frac {4\pi}{ik}\frac {\sinh k\left (\frac {\pi}{2}-\frac {\pi}{2}\beta ^2\right)}{\sinh \frac {\pi}{2}k} \cos k\Theta \, . \label {four1}
\ee
Reparametrising the coupling constant $\beta ^2$, introducing the kernel $G(\theta)$ and the renormalised twist $\tilde \omega$, 
\be
\gamma\equiv \pi(1-\beta ^2) \, , \quad \hat G(k) \equiv -\frac {\hat
  K(k)}{1-\hat K(k)}=\frac {\sinh k\left (\frac {\pi}{2}- \gamma
  \right)}{2 \cosh k \frac {\gamma}{2} \sinh k\left (\frac
    {\pi}{2}-\frac {\gamma}{2}\right)} \, , \quad \tilde \omega \equiv
\frac {\pi}{\beta ^2}\omega  \, , \label {G}
\ee
equation (\ref {count3}) takes the compact form
\be
\hat Z_N(k)=\frac {N}{4}\frac {\hat \Phi (k)}{1-\hat K(k)}+2\hat G(k)\hat L_N(k)+2 \pi  \varepsilon \tilde \omega 
{\delta (k)} \, . \label {count4}
\ee
Now we want to Fourier anti-transform and then to perform the continuous
limit: therefore, it is necessary to work out in such a limit the behaviour of  
the anti-transform of the first term in the r.h.s.
\ba
F_N(x)&\equiv& \int _{-\infty}^{+\infty} \frac {dk}{2\pi} e^{ikx}
\left (\frac {N}{4}\frac {\hat \Phi (k)}{1-\hat K(k)}\right ) =
\nonumber \\
&=&\frac {\pi N}{4i}\int _{-\infty}^{+\infty} \frac {dk}{2\pi k}
\left [ e^{ik(x+\Theta)}+e^{ik(x-\Theta)}\right ] \frac {\sinh k\left (\frac {\pi}{2}-\frac {\pi}{2}\beta ^2\right)}{\sinh k \frac {\pi}{2}\beta ^2 \cosh k\left (\frac {\pi}{2}-\frac {\pi}{2}\beta ^2\right)} \, . \label {forc}
\ea
This {\it driving term} may be evaluated through the residue theorem and it diverges in the continuous limit, unless $\Theta$
depends on $N$ in a peculiar way. We sketch here the evaluation, since it differs from the previous ones \cite {KP,DDV,FMQR} (even from the conformal case \cite {FR2}). All the poles of the integrand lie on the
imaginary axis and, when $1/3<\beta ^2< 2/3$, the poles with smallest modulus are $k=0,\pm i \frac {\pi}{\gamma}$.
If in the continuous limit
$\Theta \rightarrow +\infty$, we evaluate for fixed $x$ the integral
in (\ref {forc}) containing $e^{ik(x+\Theta)}$ by choosing a semi-circle contour
in the upper $k$-complex half-plane with a small
semicircle avoiding the pole at $k=0$. Since the contributions from
the other poles are damped by an exponential of
$\Theta$, the leading term is given by the
residue at the pole with the smallest nonzero modulus, i.e. $k=i\pi/\gamma$, minus the contribution of the semicircle around $k=0$, whose value is linear in $N$: $\frac {\pi N(1-\beta ^2)}{8\beta ^2}$. 
Analogously, in the evaluation of the integral containing $e^{ik(x-\Theta)}$
we close the contour - containing a semicircle avoiding $k=0$ - in the
lower $k$-complex half-plane. Now, the
leading contribution is given by the pole at $k=-i\pi/\gamma$, minus the contribution of the semicircle around $k=0$, whose value  
$\left (-\frac {\pi N(1-\beta ^2)}{8\beta ^2}\right)$ cancels the previous one.
Adding the two residue contributions we obtain (up to terms $o(N^0)\rightarrow 0$ when $N\rightarrow +\infty$):
\be
N\rightarrow +\infty \, , \quad F_N(x)=Ne^{-\frac {\pi}{\gamma}\Theta}\frac {\sinh \frac {\pi}{\gamma}x}{\sin \frac {\pi \beta ^2}{2(1-\beta ^2)}}+o(N^0)=mR \sinh \frac {\pi}{\gamma}x +o(N^0)\, ,
\ee
where we have supposed a logarithmic divergence of $\Theta$, via a constant mass parameter $m$ (in such a way that $mR$ is dimensionless):
\begin{equation}
\Theta = -\frac {\gamma}{\pi} \ln \left [ \frac {mR}{N} \sin \frac {\pi \beta ^2}{2(1-\beta ^2)} \right ] \, . \label {theta}
\end{equation}
With this result in mind it is easy to write an equation satisfied by the counting function in the continuous limit, $Z(x)$, from relation (\ref {count4}): \be
Z(x)=mR\sinh \frac {\pi}{\gamma}x+
2\int_{-\infty}^{+\infty}{dy}\, G(x - y){\mbox {Im}}\ln \left [1+e^{iZ(y+i0)}\right ]+ \varepsilon \tilde \omega  \, . \label{ddv} 
\ee
This {\it nonlinear integral equation} describes
the $\tilde \omega$-vacuum of the theory at the different values of $\tilde \omega$.

\medskip

{\bf Remark 1} Equation (\ref {ddv}) has been found for $1/3<\beta ^2
< 2/3$ and small $\tilde \omega$. However, it can be considered without problems in the whole
region $0<\beta ^2 <1$ and $\tilde \omega \geq 0$, defining by analytical
continuation a state of the continuous theory which minimises energy.

{\bf Remark 2} Equation (\ref {ddv}) with $\tilde \omega =0$ would also describe the vacuum of the continuous theory from the lattice sine-Gordon model \cite {IK,EFIK}, with lagrangian
\be
{\cal L}=\frac {1}{2}(\partial \varphi)^2+\frac {m_0^2}{8\pi \beta ^2}\cos
{\sqrt {8\pi}} \beta
\varphi \, , \label{SGlag}
\ee
where $m_0=2{\sqrt {2}}\left (\frac {R}{N}\right )^{-1}(\cosh 2\Theta)^{-1/2}$ is a bare mass\footnote{From (\ref {theta}) we have the renormalised mass parameter $m=\left [ \frac {m_0}{4}\left ( \frac {R}{N}\right)^{\beta ^2}\right]^{\frac {\pi}{\gamma}} \left / \sin \frac {\pi \beta ^2}{2(1-\beta ^2)}\right. $ in the continuous limit.}. 
Hence, there is a sort of interchange, given by the first of (\ref {G}), between the coupling constant $\beta $ in the lagrangian and $\gamma$ in the equation (\ref {ddv}).

\section {Local charges and transfer matrix}
\setcounter{equation}{0}
 
The continuous limit of a sum like (\ref {sumf1}) can be expressed, once (\ref {ddv}) has been put in, in terms of only the counting function $Z(x)$ (and the continuous limit of $f_N(x),f(x)$):
\be
\lim _{\stackrel {N\rightarrow +\infty}{N\Delta =R}} \sum _{r=1}^{l}f_N(\alpha _r)=f_b+\int _{-\infty}^{+\infty}\frac {dx}{\pi}
J_f(x){\mbox {Im}}\ln \left [1+e^{iZ(x+i0)}\right ]  \, . \label {sumf5}
\ee
We have rearranged the r.h.s. in order to separate the bulk contribution 
\be
f_b=\lim _{N\rightarrow +\infty}\left [-\int _{-\infty}^{+\infty}\frac {dx}{2\pi} f^\prime _N(x)
\left ( mR \sinh \frac {\pi}{\gamma} x
+\varepsilon \tilde \omega  \right ) \right ]\, ,  \label {bulk}
\ee
from the finite size corrections depending on the linear functional
\be
J_f(x)=\lim _{N\rightarrow +\infty} \int _{-\infty}^{+\infty}\frac {dk}{2\pi}e^{ikx} ik \hat f_N(k) \hat J(k) \, , \quad
\hat J(k)=\frac {\sinh k\frac {\pi}{2}}{2 \cosh k \frac {\gamma}{2} \sinh k\left (\frac {\pi}{2}-\frac {\gamma}{2}\right)} \, . \label {J}
\ee
Formula (\ref {sumf5}) allows to calculate exactly state depending quantities of a Quantum Field Theory on cylinder. 
In particular this formula applies effectively to the calculation of the energy of the $\tilde \omega$-vacuum. We build up the hamiltonian operator in tight analogy with the same construction in lattice sine-Gordon model
\begin{eqnarray}
H&\equiv&-\frac {1}{4\Delta \sin \pi \beta ^2 \cosh 2 \Theta } \Bigl
[e^{-2\alpha} \frac {\partial}{\partial \alpha}G^-_N (\alpha)\left
  |_{\alpha=-\Theta -\frac {i\pi \beta ^2}{2}}\right. +e^{-2\alpha}
\frac {\partial}{\partial \alpha}G^+_N(\alpha)\left |_ {\alpha=-\Theta
    +\frac {i\pi \beta ^2}{2}}\right.  \nonumber \\ 
 &-&e^{2\alpha}\frac {\partial}
{\partial \alpha}G^-_N(\alpha)\left |_{\alpha=\Theta -\frac {i\pi \beta ^2}{2}}\right.-e^{2\alpha}\frac {\partial}{\partial
  \alpha }G^+_N(\alpha)\left |_{\alpha=\Theta +\frac {i\pi \beta
      ^2}{2}}\right.\Bigr ] \, ,   \label {Ham}
\end{eqnarray}
where we have defined $G^{\pm}_N(\alpha)\equiv\ln \left [\rho _N^{\pm}(\alpha)^{-1}{\bf \Lambda }_N(\alpha)\right ]$.
Using (\ref {tra}-\ref {rho}, \ref {Ham}), the eigenvalues of $H$ for a general Bethe state read as a sum on Bethe roots, $\sum \limits _{r=1}^l h(\alpha _r,N,R)$,
where
\ba
h(x,N,R)=-\frac {1}{4\Delta \sin \pi \beta ^2 \cosh 2 \Theta }\Bigl [ -\frac {e^{\Theta -x -\frac {i\pi \beta ^2}{2}}}{\sinh \left (x+\Theta +\frac {3i\pi \beta ^2}{2}\right )} + \frac {e^{\Theta -x +\frac {i\pi \beta ^2}{2}}}{\sinh \left (x +\Theta +\frac {i\pi \beta ^2}{2} \right )}  &&\nonumber \\
+\frac {e^{3\Theta -x-\frac {5i\pi \beta ^2}{2}}}{\sinh \left (x -\Theta +\frac {3i\pi \beta ^2}{2} \right )}  
 -\frac {e^{3\Theta -x -\frac {3i\pi \beta ^2}{2}}}{\sinh \left (x -\Theta +\frac {i\pi \beta ^2}{2} \right )}  
 -\frac {e^{\Theta -x +\frac {i\pi \beta ^2}{2}}}{\sinh \left (x +\Theta -\frac {3i\pi \beta ^2}{2} \right )} &\, & \label {en1} \\
 +\frac {e^{\Theta -x-\frac {i\pi \beta ^2}{2}}}{\sinh \left (x +\Theta -\frac {i\pi \beta ^2}{2} \right )}  
 +\frac {e^{3\Theta -x +\frac {5i\pi \beta ^2}{2}}}{\sinh \left (x -\Theta -\frac {3i\pi \beta ^2}{2} \right )}  
 -\frac {e^{3\Theta -x +\frac {3i\pi \beta ^2}{2}}}{\sinh \left (x -\Theta -\frac {i\pi \beta ^2}{2} \right )} \Bigr ] \, .&& \nonumber
\ea
When $1/3<\beta ^2<2/3$, extending the results of \cite
{BOG} to $\tilde \omega \not= 0$, we are able to show that the $\tilde \omega$-vacuum is indeed the eigenstate with the lowest eigenvalue.
Moreover, when the twist $\tilde \omega =0$ this is the energy of the lattice sine-Gordon model \cite {IK}.

After Fourier transforming \cite {GR}
\ba
ik\hat h(k,N,R)&=&\frac {ie^{2\Theta}}{\Delta \cosh 2\Theta \sin \pi \beta ^2}\frac {\pi k \sinh k\frac {\pi}{2}\beta ^2}{\sinh k\frac {\pi}{2}}\Bigl [ \sin \left (k\Theta -\pi \beta ^2\right) e^{-k\frac {\pi}{2}+k\pi \beta ^2}-\nonumber \\
&-&  \sin \left (k\Theta +\pi \beta ^2\right ) e^{k\frac {\pi}{2}-k\pi \beta ^2} \Bigr ] \, , \label {en3}
\ea
we are left with the evaluation of $J_h(x)$ (\ref {J}) in the continuous limit
\ba
J_h(x)&=&\lim _{N\rightarrow \infty} \left \{ \frac {Nie^{2\Theta}}{2 R \cosh 2\Theta \sin \pi \beta ^2}
\int _{-\infty}^{+\infty} \frac {dk}{2\pi}e^{ikx}\frac {\pi k}{\cosh
  k\frac {\gamma}{2}}\Bigl [ \sin \left (k\Theta -\pi \beta ^2\right )
e^{-k\frac {\pi}{2}+k\pi \beta ^2} - \right. \nonumber \\
&-&\left. \sin \left (k\Theta +\pi \beta ^2\right ) e^{k\frac {\pi}{2}-k\pi \beta ^2} \Bigr ] \right \} \, . \label {I2}
\ea
Again, because of relation (\ref {theta})
we can calculate exactly this limit by using the residue method (closing the contour of integration in the upper or in the lower $k$-complex half-plane).
Collecting the finite contributions we obtain a simple behaviour
\be
J_h(x)=- 4m\frac {\sin \left [\frac {\pi}{2} \frac {1-2\beta ^4}{1-\beta ^2} \right] \sin \frac {\pi \beta ^2}{2(1-\beta ^2)}}{(1-\beta ^2)^2 \sin \pi \beta ^2} \sinh \frac {\pi}{\gamma}x 
\, . \label {I3}
\ee
Hence, the finite size correction $E-E_b$
to the energy of the $\tilde \omega$-vacuum is 
\be
E-E_b=-4m \frac {\sin \left [\frac {\pi}{2} \frac {1-2\beta ^4}{1-\beta ^2} \right] \sin \frac {\pi \beta ^2}{2(1-\beta ^2)}}{(1-\beta ^2)^2 \sin \pi \beta ^2}
\int _{-\infty}^{+\infty}\frac {dx} {\pi} \sinh \frac {\pi}{\gamma}x
\, {\mbox {Im}}\ln \left [1+e^{iZ(x+i0)}\right ] \, . \label{EN}
\ee
As far as the $\tilde \omega$-vacuum state is concerned, this formula has been derived within a twisted lattice sine-Gordon model and hence proves the quantum equivalence with the approaches to the sine-Gordon model based on spin $+1/2$ XXZ chain \cite {DDV,FMQR}. However, a different renormalisation of the mass occurs in (\ref {EN}).   

\medskip  

We now want to compute the continuous limit of the
eigenvalues of the transfer matrix (\ref {tra}) proposed in \cite {FR} as
description of perturbed conformal field theories. 
To make formula (\ref {sumf5}) useful, we shall define  
${\bf F}^{\pm}_N(\alpha )\equiv\ln {\bf \Lambda}^{\pm}_N(\alpha )$. Let us firstly concentrate on ${\bf F}^+_N(\alpha)$:
\ba
&&{\bf F}^+_N(\alpha )=\sum _{r=1}^{l}\Bigl [ \ln \frac {\sinh (\alpha -\alpha _r +i\pi \beta ^2)}{\sinh (\alpha -\alpha _r)}  -i\pi \varepsilon \beta ^2\Bigr]+\nonumber \\
&+&\frac {N}{4}\left [ \ln 2 \sinh \left ( \Theta -\alpha -\frac {i\pi \beta ^2}{2} \right ) + \ln 2 \sinh \left ( \Theta +\alpha +\frac {i\pi \beta ^2}{2} \right ) -2\Theta \right ]  \, . \label {tra3}
\ea
Now we restrict ourselves to the $\tilde \omega$-vacua, although generalisations to excited states can be worked out along the developments of \cite {FMQR}.
The last addendum in (\ref {tra3}) gives a contribution whose
continuous limit is zero if $0<\beta ^2 <1/2$ and infinity if
$1/2<\beta ^2 <1$. Therefore, we shall regularise ${\bf F}^+(\alpha)$ for
$1/2<\beta ^2 <1$ defining it as the analytic continuation of ${\bf F}^+(\alpha)$ with $0<\beta ^2 <1/2$. With this regularisation in mind we obtain from (\ref {sumf5}) 
\be
{\bf F}^+(\alpha )={\bf F}_b^+(\alpha )+\int _{-\infty}^{+\infty}\frac {dx}{\pi}
\Bigl [\int _{-\infty}^{+\infty}\frac {dk}{2\pi}e^{ikx} ik \hat f^+(k,\alpha) \hat J(k) \Bigr ]{\mbox {Im}}\ln \left [1+e^{i Z(x+i0)}\right ] -\varepsilon i\pi \omega \, , \label {sumf6}
\ee
where ${\bf F}_b^+(\alpha)$ is the bulk contribution, $\hat J(k)$ is given by (\ref {J}) and 
\be
f^+(x,\alpha )=\ln \frac {\sinh \left (\alpha -x+i\pi \beta ^2 \right)}{\sinh (\alpha -x)}  \, . \label {f3}
\ee
The twist $\omega$ comes out as in \cite {FR2}.
Since ${\bf F}^+(\alpha )$ (\ref {tra3}) has the periodicity property
${\bf F}^+(\alpha +i\pi)={\bf F}^+(\alpha)$, it is sufficient to study ${\bf F}^+(\alpha)$ in the strip
$\gamma -\pi <{\mbox {Im}}\alpha <\gamma$. 
Comparing (\ref {sumf6}) with relation (5.6) of \cite {FR2}, we remark
that the second term in the r.h.s. is formally the
same. Therefore, we can use the results of that paper:
from formul{\ae} (5.16, 5.24) of \cite {FR2} we obtain that, if $0<{\mbox
  {Im}}\alpha <\gamma $, 
\be
{\bf F}^+(\alpha)={\bf F}_b^+(\alpha)-\int _{-\infty}^{+\infty}\frac {dx}{\gamma}\frac {1}{\sinh \frac {\pi}{\gamma}(x-\alpha)}{\mbox {Im}}\ln \left [1+e^{iZ(x+i0)}\right ] -\varepsilon i\pi \omega\, ,
\label {sumf7}
\ee
and that, if $\gamma -\pi< {\mbox {Im}}\alpha < 0$,
\ba
&&{\bf F}^+(\alpha)={\bf F}_b^+(\alpha)-\varepsilon i\pi \omega +
\label{resumf7} \\
&+&\int _{-\infty}^{+\infty}\frac {dx}{\pi }\left \{ \int _{-\infty}^{+\infty}
{dk}e^{ik\left (x-\alpha-\frac {i\pi\beta ^2}{2}\right )} \frac {i\sinh k \frac {\gamma}{2}}{2\cosh k\frac {\gamma}{2}\sinh k\left (\frac {\pi}{2}-\frac {\gamma}{2}\right)}\right \}{\mbox {Im}}\ln \left [1+e^{iZ(x+i0)}\right ] \, .
\nonumber
\ea
On the other hand, from definitions (\ref {bulk}) and (\ref {f3}) it follows that
\be
{\bf F}_b^+(\alpha)=\int _{-\infty}^{+\infty}\frac {dx}{2\pi}
\frac {\sinh i\pi \beta ^2}{\cosh \left (\alpha-x-\frac {i\pi}{2}\right ) \cosh \left (\alpha-x-\frac {i\pi}{2}+i\pi\beta^2\right )}\left [mR\sinh \frac {\pi}{\gamma} x
+\varepsilon \tilde \omega \right ]\, . \label {bulk3}
\ee
This integral is finite only for $0<\beta ^2 < 1/2$. If $0<{\mbox {Im}}\alpha<\gamma$ its value is
\be
{\bf F}^+_b(\alpha)=-mR \, {\mbox {ctg}} \, \frac {\pi ^2}{2\gamma}
\cosh \frac {\pi}{\gamma}\alpha+i\varepsilon \pi \omega \, , \label {bulk4}
\ee
whereas, if $\gamma-\pi< {\mbox {Im}}\alpha < 0$ it equals 
\be
{\bf F}_b^+(\alpha)=-mR \, \frac {1}{\sin \frac {\pi^2 }{2\gamma}}
\cosh \left [ \frac {\pi}{\gamma}\left (\alpha+i\frac {\pi}{2}\right )\right ] +i\varepsilon \pi \omega -i\varepsilon \tilde \omega \, . \label {rebulk4}
\ee

Let us now consider ${\bf F}^-(\alpha)$ which reads from (\ref {tra}-\ref {rho}) \ba
&&{\bf F}^-_N(\alpha )=\sum _{r=1}^{l}\Bigl [ \ln \frac {\sinh (\alpha -\alpha _r -i\pi \beta ^2)}{\sinh (\alpha -\alpha _r)}  +i\pi \varepsilon \beta ^2 \Bigr]+\nonumber \\
&+&\frac {N}{4}\left [ \ln 2 \sinh \left ( \Theta -\alpha +\frac {i\pi \beta ^2}{2} \right ) + \ln 2 \sinh \left ( \Theta +\alpha -\frac {i\pi \beta ^2}{2} \right ) -2\Theta \right ]  \, . \label {2tra3}
\ea
Again, in the continuous limit the last term is zero
in the interval $0<\beta ^2<1/2$ and we use our regularisation
procedure. According to the latter we obtain a relation
\be
{\bf F}^-(\alpha )=-{\bf F}^+\left (\alpha-i\pi \beta ^2 \right ) \, , \label{f2f1}
\ee
which gives ${\bf F}^-(\alpha)$ for any complex $\alpha$. From the
knowledge of ${\bf F}^{\pm}(\alpha)$ it is simple to reconstruct
${\bf \Lambda}(\alpha)={\mbox {exp}}[{\bf F}^+(\alpha)]+{\mbox
  {exp}}[{\bf F}^-(\alpha)]$ in all the
$\alpha$-complex plane.   

From the classical results \cite {FS} we can argue that the coefficients of the asymptotic expansion of $\ln {\bf \Lambda }(\alpha)$ are proportional to the local
integrals of motion in involution ${\bf I}_{2n+1}^{\mp}$ of perturbed conformal field 
theories. Therefore, using
the same techniques as in \cite {FR2}, we expand $\ln {\bf
  \Lambda} (\alpha)$ for ${\mbox {Re}} \alpha \rightarrow \pm \infty$
in the strip ${\mbox {max}}\left \{-\frac {\pi}{2},-\gamma\right\}<{\mbox {Im}} \alpha < {\mbox {min}}\left\{\frac {\pi}{2},\gamma\right\}$ 
(which is only a technical assumption):
\ba
&&\ln {\bf \Lambda} (\alpha)\, \, {\stackrel {\cdot}{=}}\, \,   -
mR\, {\mbox {ctg}} \, \frac {\pi ^2}{2\gamma}\cosh \frac {\pi}{\gamma}\alpha
\pm \nonumber \\
&\pm&\frac {2}{\gamma}\sum _{n=0}^{+\infty}e^{\mp \frac
  {\pi}{\gamma}\alpha (2n+1)}
\int _{-\infty}^{+\infty} dx \,  e^{\pm\frac {\pi}{\gamma}x(2n+1)}\, {\mbox {Im}}\ln \left [1+e^{iZ(x+i0)}\right ] \, . \label{f1cont}
\ea
Now, we need to compare this expression with the conformal analogous (5.48) of \cite {FR2} and to consider the conformal normalisation coefficients $c_n$ ((5.50) of \cite {FR2}), to write down
\be
\ln {\bf \Lambda} (\alpha)\, \, {\stackrel {\cdot}{=}}\, \,   -
mR \, {\mbox {ctg}} \, \frac {\pi ^2}{2\gamma}\cosh \frac {\pi}{\gamma}\alpha
\pm \frac {2}{\pi}\sum _{n=0}^{+\infty}e^{\mp \frac
  {\pi}{\gamma}\alpha (2n+1)}c_{n+1} {\bf I}_{2n+1}^{\mp} \, , \label{expa}
\ee
where the local integrals of motion have the exact expressions ($\chi =\mp1$):
\be
 {\bf I}_{2n+1}^{\chi}=\frac {1}{c_{n+1}}\frac {\pi}{\gamma}
\int _{-\infty}^{+\infty} dx \,  e^{-\chi \frac {\pi}{\gamma}x(2n+1)}\, {\mbox {Im}}\ln \left [1+e^{iZ(x+i0)}\right ] \, . \label {In}
\ee
Indeed, we have renormalised them with respect to the conformal counterparts $I_{2n+1}$ (see (5.51) of \cite {FR2}) so that in the conformal limit
\be
\lim _{r \rightarrow 0}\left (\frac {r}{2A}\right)^{2n+1} {\bf I}_{2n+1}^{\chi =\varepsilon}=I_{2n+1}
\, , \label{connection}
\ee
where $r=mR$ is a dimensionless parameter.
Moreover, we have been able to build up a lattice field theory for which conjecture (3.31) of \cite {BLZ2} - suggested by pure analogy with the conformal case \cite {BLZ} - holds. Nevertheless, the na\"{\i}ve scaling limit of our monodromy matrix is absolutely different from the proposed (but not actually used) prescription given in \cite {BLZ2}. On the other hand, we are in agreement with the finding of \cite {BLZ2} derived using a brilliant development of the Thermodynamic Bethe Ansatz (TBA) technique in the case of Lee-Yang model \cite {CMZ}.

For $\beta ^2 =1/2$ the local integrals of motion 
(\ref {In}) are explicitly given by 
\be
{\bf I}_{2n+1}^{\chi}=-2^{1-n}\pi^{2n}(n+1)(2n+1){\mbox {Im}} \int _{-\infty}^{+\infty} dx 
e^{-\chi 2(x+i\eta)(2n+1)}\ln \left \{ 1+e^{i[mR\sinh 2(x+i\eta)+2\pi\varepsilon \omega]}\right \}, \label{In3}
\ee
where $0<\eta<\pi/2$. The choice $\eta =\pi/4$ gives 
\be
{\bf I}_{2n+1}^{\chi}=\chi (-1)^{n}2^{1-n}\pi ^{2n}(n+1)(2n+1) r^{-2n-1} \, {\mbox {Re}} P_{2n+1} (a, r) \, ,
\ee
where $a =-2\pi  \varepsilon (\omega +1/2)$ and we have introduced the rescaled charges 
\be
P_{2n+1}(a, r)=r^{2n+1}\int _{0}^{+\infty}d\theta \, \cosh \theta (2n+1)\,  \ln \left [ 1-e^{-r\cosh \theta -ia}\right] \, ,   \label {Q1}
\ee
which have a finite conformal limit ($r\rightarrow 0$). Their derivative with respect to $r$ is written, after an integration by parts, in terms of modified Bessel functions,
\be
\frac {dP_{2n+1}}{dr}(a, r)=r^{2n+1}\sum _{k=1}^{+\infty}e^{-ia k} K_{2n}(kr) \, . \label{P2}
\ee
After integrating this formula and considering that the integration constant is fixed by the known \cite {FR2} conformal limit (\ref {connection}), we would obtain an expression of ${\bf I}^{\chi}_{2n+1}$ as series of modified Bessel functions. Nevertheless, we prefere to differentiate (\ref {P2}) and write down a relation involving $K_0$ only \cite {GR}:
\be
\left (\frac {1}{r} \frac {d}{dr} \right )^{2n+1}{\mbox {Re}}P_{2n+1}(a, r) =\sum _{k=1}^{+\infty}\cos{ak} \, k^{2n} K_{0}(kr)=(-1)^n\frac {d^{2n}}{{da}^{2n}} \sum _{k=1}^{+\infty}\cos{ak} K_{0}(kr)\, . \label {P3}
\ee
This expression allows us to simplify the form of the off-critical charges via a Schl\"{o}milch formula \cite {GR} (${\bf C}$ is the Euler-Mascheroni constant):
\ba
&&\sum _{k=1}^{\infty} K_0(kr) \cos ka=\frac {1}{2}\left ({\bf C}+\ln \frac {r}{4\pi}\right ) +\frac {\pi}{2{\sqrt {r^2+a^2}}} +\nonumber \\
&+& \frac {\pi}{2} \sum _{l=1}^{\infty} \left [ \frac {1}{{\sqrt {r^2+(2l\pi -a)^2}}}+
 \frac {1}{{\sqrt {r^2+(2l\pi +a)^2}}}-\frac {1}{\pi l} \right ] \, . \label{Sch}
\ea
Term by term derivatives can be performed using the formula
\be
\frac {d^{2n}}{{da}^{2n}}\frac {1}{{\sqrt {x^2+a^2}}}=\frac {\sum \limits _{\stackrel {h=0}{h\in 2{\mathbb N}}}^{2n}b_{h}^{(2n)}a^{h}x^{2n-h}}{(x^2+a^2)^{2n+\frac {1}{2}}} \, , \quad 
b_h^{(2n)}= 
\frac {(-1)^{n+\frac {h}{2}}\left [ (2n-1)!!\right ] ^2}{h!}\prod \limits_{\stackrel {k=0}{k\in 2{\mathbb N}}}^{h-2}(2n-k)^2  \, .
\ee
We give the explicit result for the case $n=0$,
\ba
&&{\mbox {Re}}P_{1}(a, r)=\frac {r^2}{8} [ 2{\bf C}-2 \ln 4\pi +\ln r^2 - 1]+\frac {\pi}{2} {\sqrt {r^2+a^2}}-\frac {\pi}{2}|a|+\label{P5} \\
&+&\frac {\pi}{2}\sum _{l=1}^{\infty}\left [ {\sqrt {r^2+(2l\pi -a)^2}} +
{\sqrt {r^2+(2l\pi +a)^2}} -\frac {r^2}{2\pi l}-4\pi l \right]
-\pi ^2 B_2\left (\frac {a}{2\pi}\right) \, , \nonumber 
\ea 
because in the cases $\omega =0, \pm 1/2$ it reproduces known TBA results \cite {KM}.

So far, we have studied the monodromies of ${\bf I}_{2n+1}^{\chi}$ on $r=mR$.
Actually, it is also important to see how the charges depend on the twist $\omega$.
Let us extend (\ref {In3}) to semi-integers $n\geq 0$. Introducing the index $k\in {\mathbb N}$, $k \geq 1$,
we find that the following differential equation holds:
\be
\left (\frac {r\chi}{2\pi \varepsilon}\right)\frac {1}{2^{1/2}\pi (k+3)}\frac {d {\bf I}^{\chi}_{k+2}}{d\omega}+ \left(\frac {r\chi}{2\pi \varepsilon}\right)\frac {\pi (k+2)}{2^{3/2}k(k+1)}\frac {d {\bf I}^{\chi}_{k}}{d\omega}-{\bf I}^{\chi}_{k+1}=0
\,  . \label{eqdif}
\ee
In the conformal limit $r\rightarrow 0$, taking into account (\ref {connection}), we get the solution $I_{k+1}=-\varepsilon 2^{-1-k/2}B_{k+2}(\omega +1/2)$.

\section {Summary and outlook}

Starting from the lattice theory of two coupled (m)KdV equations, we 
have found in the whole interval
$0<\beta ^2 <1$ the nonlinear integral equation which describes the
vacuum sector and we are in the position to work out all features of excited states after \cite {FMQR}. The existence in our approach of a
transfer matrix (which generates the local integrals of motion in
involution) has allowed us to give expressions for its
eigenvalues on the twisted vacua in terms of the
solutions of the nonlinear integral equation and makes possible, in the next future, to find a deep link with the Thermodynamic Bethe Ansatz technique of \cite {BLZ2}. We have found strong evidence
that local charge eigenvalues coincide with the primary state eigenvalues of the
local integrals of motion of $\Phi _{1,3}$ perturbed minimal conformal models, the primary weight being linked to the twist. The expressions become explicit for $\beta ^2=1/2$.
Eventually, we want to emphasise that the wide generality of our construction \cite {FR,FR2} makes possible its application in the domain of two-dimensional supersymmetric field theories.

{\bf Acknowledgements} - Discussions with R. Tateo and
P. Dorey are kindly acknowledged. D.F. thanks EPRSC (grant GR/M66370)
and Leverhulme Trust (grant F/00224/G). M.R. thanks EPRSC for the
grant GR/M97497 and the Department of Mathematical Sciences of Durham
for warm hospitality. This work has also been supported by EC FP5
Network, contract number HPRN-CT-2002-00325.

\end{document}